\documentclass[twocolumn,showpacs,superscriptaddress,aps,prl]{revtex4-1}
\usepackage{graphicx}
\usepackage[caption=false,subrefformat=parens,labelformat=parens]{subfig}
\usepackage{hyperref}
\hypersetup{
    colorlinks=true,       		
    linkcolor=blue,        
    citecolor=blue,        
    filecolor=blue,      	
    urlcolor=blue         	
}

\newcommand{\bQ}{\mbox{\boldmath$Q$}}

\begin{document}

\title{Spin liquid polymorphism in a correlated electron system on the threshold of superconductivity}

\author{Igor A.\ Zaliznyak}
\email{zaliznyak@bnl.gov}
\affiliation{CMPMSD, 
 Brookhaven National Laboratory, Upton, NY 11973 USA}

\author{Andrei Savici}
\affiliation{QCMD, 
 Oak Ridge National Laboratory, Oak Ridge, TN 37831 USA}

\author{Mark Lumsden}
\affiliation{QCMD, 
 Oak Ridge National Laboratory, Oak Ridge, TN 37831 USA}

\author{Alexei Tsvelik}
\affiliation{CMPMSD, 
 Brookhaven National Laboratory, Upton, NY 11973 USA}

\author{Rongwei Hu}
\affiliation{CMPMSD, 
 Brookhaven National Laboratory, Upton, NY 11973 USA}
\affiliation{Present address: Rutgers Center for Emergent Materials and Department of Physics and Astronomy, Rutgers University, Piscataway, New Jersey 08854, USA}

\author{Cedomir Petrovic}
\affiliation{CMPMSD, 
 Brookhaven National Laboratory, Upton, NY 11973 USA}



\begin{abstract}
{We report neutron scattering measurements, which reveal spin-liquid polymorphism in a '11' iron chalcogenide superconductor, a poorly-metallic magnetic FeTe tuned towards superconductivity by substitution of a small amount of Tellurium with iso-electronic Sulphur. We observe liquid-like magnetic dynamics, which is described by a competition of two phases with different local structure, whose relative abundance depends on temperature.
One is the ferromagnetic (FM) plaquette phase observed in the non-superconducting FeTe, which preserves the C$_4$ symmetry of the underlying square lattice and is favored at high temperatures. The other is the antiferromagnetic plaquette phase with broken C$_4$ symmetry, which emerges with doping and is predominant at low temperatures.
These findings suggest a first-order liquid-liquid phase transition in the electronic spin system of FeTe$_{1-x}$(S,Se)$_x$. We thus discover remarkable new physics of competing spin liquid polymorphs in a correlated electron system approaching superconductivity. Our results facilitate an understanding of large swaths of recent experimental data in unconventional superconductors.}
\end{abstract}


\maketitle

Ideas about liquid polymorphism date as far back as Mendeleev's work on water-alcohol mixtures and R\"{o}ntgen's proposals that many of the anomalous properties of water, such as its density maximum at 4$^\circ$C, can be understood if water is viewed as a mixture of two liquid "species", different in local structure and density \cite{Mendeleev1887,Rontgen1891,Pauling1939}. The possible existence of different liquid polymorphs and liquid-liquid phase transitions (LLPT) between such polymorphs in simple molecular fluids continues to receive considerable attention, but the issue remains unsettled \cite{Poole_etal_Science1997,Poole_etal_Nature1992,MishimaStanley_Nature1998,Mallamace_PNAS2009,Huang_etal_PNAS2009,Clark_etal_PNAS2010,Beye_etal_PNAS2010}. This is mainly because the competition between different liquid phases is governed by the energy balance of inter-molecular interactions, and is thus expected only at low temperatures, where it outweighs thermal entropy. For most simple liquids this is expected only in a deeply supercooled region, well below the freezing temperature \cite{Poole_etal_Science1997,Poole_etal_Nature1992,MishimaStanley_Nature1998,Mallamace_PNAS2009,Huang_etal_PNAS2009,Clark_etal_PNAS2010,Beye_etal_PNAS2010}. On the other hand, such competition arises quite naturally in a system of atomic magnetic moments in a crystal, where development of magnetic order is hindered by competing interactions (frustration) \cite{LeeBroholm_etal_Nature2002,StoneZaliznyak_etal_Nature2005}, reduced dimensionality \cite{Zaliznyak_NatureMater2005,Walters_etal_NaturePhys2009}, or both \cite{Han_etal_Nature2012}. In these cases, the system remains disordered even at temperatures much lower than the energy of magnetic interactions, thus realizing a spin-liquid state. Hence, many exotic quantum properties of liquids have been discovered by studying spin liquids \cite{LeeBroholm_etal_Nature2002,StoneZaliznyak_etal_Nature2005,Zaliznyak_NatureMater2005,Walters_etal_NaturePhys2009,Han_etal_Nature2012}.

Liquid-like states play a prominent role in theories of unconventional superconductivity in cupric oxides and iron pnictides and chalcogenides, which feature two-dimensional square nets of magnetic atoms with a strongly reduced tendency to order magnetically \cite{Anderson_Science1987,LeeNagaosaWen_RMP2006,Scalapino_RMP2012}. There is strong experimental evidence that superconductivity in these materials is coupled with the antiferromagnetic correlations, which could be responsible for electron pairing \cite{Scalapino_RMP2012}. A simple pedestrian way to visualize such coupling is to imagine that an electron moving in a spin-liquid background, by virtue of the hybridization of its wave function with those of magnetic electrons, is "dressed" by a magnetic polaron, and that this magnetic polarization cloud plays similar role to polarization of atomic displacements in the phonon mechanism of Cooper pairing in conventional superconductors. The presence of delocalized electrons interacting with the system of atomic spins also affects the nature of spin liquid state: the emerging patterns of short-range magnetic correlation not only have to optimize the orbital overlap energy of localized valence electrons (spin superexchange) but also the hybridization energy with the delocalized electron.

Motivated to uncover magnetic signatures of electronic pairing interactions, we carried out neutron scattering study of dynamical magnetic correlations in FeTe$_{0.87}$S$_{0.13}$, an iron chalcogenide of "11" superconductor family, where substitution of only 13\% of Te with S, far below any percolation phenomena, leads to the onset of superconductivity \cite{Hu_etal_PRB2009}. Through neutron scattering measurements, we have discovered not only the hallmark of spin liquid behavior, but also clear evidence of LLPT between two spin liquid polymorphs, one featuring FM plaquettes present in the parent Fe$_{1+y}$Te compound \cite{Zaliznyak_etal_PRL2011}, and the new one, which can be associated with the emergent superconductivity in FeTe$_{1-x}$(S,Se)$_x$ materials \cite{Liu_etal_NatureMater2010,Lumsden_etal_NaturePhys2010,Argyriou_etal_JPCM2010,Babkevich_etal_JPCM2012,Li_etal_PRL2010,Chi_etal_PRB2011,Babkevich_etal_PRB2011,Prokes_etal_PRB2012,Tsyrulin_etal_NJP2012,Xu_etal_PRL2012,Wen_etal_PRB2013} (see also Supporting Information).

\begin{figure}[t!h]
\centerline{\includegraphics[width=.4\textwidth]{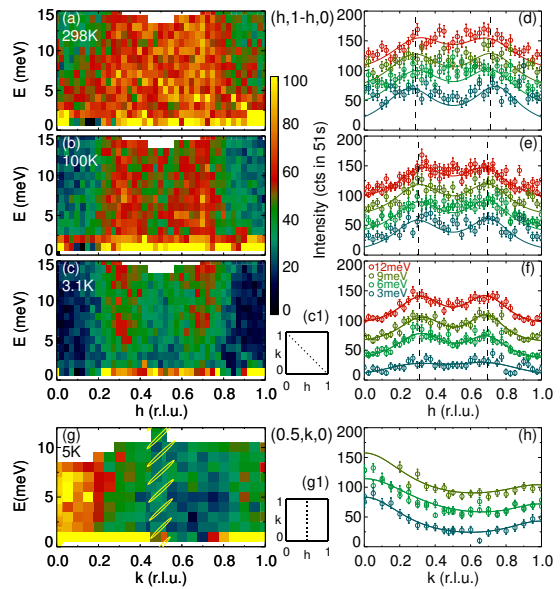}}
\caption{Dynamical magnetic correlations in FeTe$_{0.87}$S$_{0.13}$ measured by inelastic neutron scattering. (a)--(c) $(\bQ,E)$ intensity maps along $(h,1-h,0)$ direction in the $a-b$ plane, (c1), measured at 298 K, 100 K and 3.1 K, respectively. The corresponding (d)--(f) panels on the right show representative constant-E linear scans with resolution-corrected Lorentzian fits; the intensity is offset by adding (E-3)*30 counts to avoid overlaps. Dashed lines show the average peak positions. (g) $(\bQ,E)$ intensity map, and (h) representative scans along $(0.5,k,0)$ direction, (g1), at T = 5 K. Ellipses show the full width at half maximum (FWHM) of the instrument resolution.}\label{Fig1_hhEmap}
\end{figure}

\begin{figure}
\centerline{\includegraphics[width=.4\textwidth]{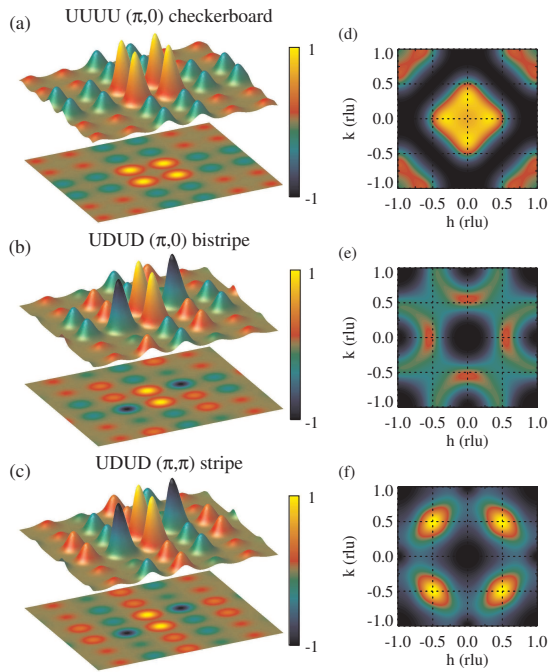}}
\caption{Electronic density and magnetization correlation in plaquette spin liquid models. (a)--(c) Height shows correlated electronic density corresponding to the isotropic atomic magnetic form factor of Fe$^{2+}$ and enlarged by the covalence factor of 2, which best fits the data in Figure \ref{Fig3_hk0map6meV}. Color represents the amplitude, in arbitrary units, of magnetic correlation for the inter-plaquette correlation length equal to two nearest-neighbour Fe-Fe spacings. A damped-wave correlation of four-iron ferromagnetic (FM) UUUU plaquettes, propagating with the wave vector $(\pi,0)$, gives rise to short-range checkerboard in FeTe, (a). For antiferromagnetic UDUD plaquettes emergent in FeTe$_{1-x}$(S,Se)$_x$, a correlation with wave vector $(\pi,0)$ gives short-range bistripe, (b), while for wave vector $(\pi,\pi)$ it gives collinear antiferromagnetism observed in iron pnictides, (c). (d)--(f) show pattern of scattering intensity corresponding to the model on the left, averaged over possible plaquette orientations to restore the macroscopic C$_4$ symmetry.}\label{Fig2_Plaquettes}
\end{figure}

\begin{figure}
\centerline{\includegraphics[width=.4\textwidth]{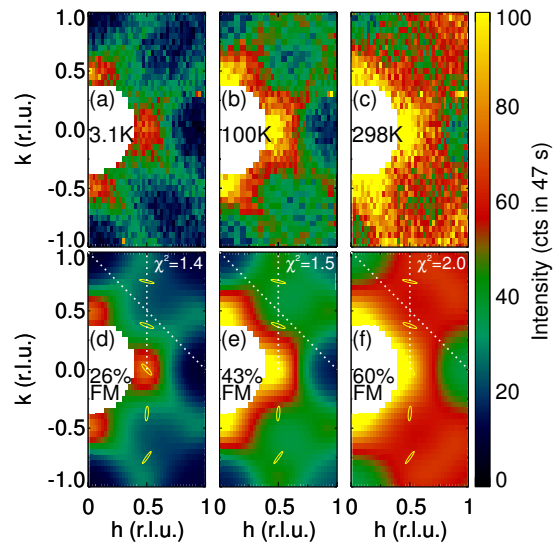}}
\caption{Neutron scattering pattern of dynamical magnetic correlations in the $a-b$ plane at constant energy E = 6 meV. (a)--(c) data at T = 3.1 K, 100 K and 298 K, respectively, as a function of the in-plane wave vector $(h,k,0)$. (d)--(e) are fits of the above data to the model of a mixture of two types of plaquette spin liquids shown in Figure \ref{Fig2_Plaquettes}. The calculated intensity is orientation-averaged to restore the macroscopic C$_4$ symmetry observed in experiment. In addition to the background and the intensity scale factor, the inter-plaquette correlation length, $\xi$, covalent compression of the magnetic form factor, and the percentage fraction of the ferromagnetic UUUU plaquettes were varied in these fits. Ellipses show the FWHM of the instrument resolution; dotted lines show locations of cuts presented in Figure \ref{Fig1_hhEmap}.}\label{Fig3_hk0map6meV}
\end{figure}

\begin{figure}
\centerline{\includegraphics[width=.4\textwidth]{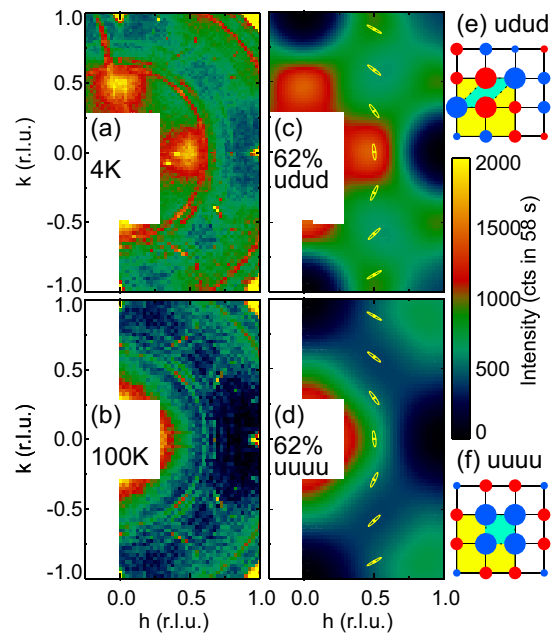}}
\caption{Quasielastic magnetic correlations measured at E = 0, showing most clearly change in the nature of dominant magnetic plaquette with temperature. (a), (b) elastic intensity at 4 K and 100 K, respectively. Data in (b) was measured for $k < 0$ and then symmetrized with respect to $k = 0$. (c), (d) show fits of the corresponding data on the left to the model of a mixture of UDUD plaquettes, (e), prevailing at low temperature, and the UUUU (FM) plaquettes, (f), emergent on warming.}\label{Fig4_hk0map0meV}
\end{figure}

The FeTe$_{1-x}$S$_x$ ($x = 0.13(1)$, m = 2.6 g) crystal used in our present measurements was grown from Te-S self-flux using a high-temperature flux method \cite{Hu_etal_PRB2009} and had a mosaic of $< 1^\circ$ full width at half maximum (FWHM). Elemental composition, structure and bulk properties, including the superconducting transition, of a small piece of the same crystal were characterized in \cite{Hu_etal_PRB2009}. Sample was mounted on an aluminium holder attached to the cold head of a closed-cycle refrigerator on the HB3 triple axis spectrometer at the High Flux Isotope Reactor at Oak Ridge National Laboratory, USA. Measurements in Figures \ref{Fig1_hhEmap} through \ref{Fig4_hk0map0meV} were performed in $(h, k, 0)$ zone with the crystal $c-$axis vertical; measurements in the $(h, 0, l)$ zone are presented in the Supporting Information. Constant-energy scans were acquired by varying the wave vector transfer  for fixed monitor counts in a low-efficiency detector between the pyrolytic graphite (PG (002)) monochromator and the sample and with fixed scattered neutron final energy E$_f$ = 14.7 meV selected by PG(002) analyzer. Beam angular collimation was set to $48'-60'-80'-240'$, from monochromator to detector, and a polycrystalline PG filter after the sample was used to minimize intensity at harmonics of the desired wavelength. The resultant energy resolution measured by the FWHM of the incoherent elastic scattering signal was 1.35 meV. The wave vector resolution calculated using standard methods (Supplementary Information) is shown by FWHM ellipses in Figures \ref{Fig1_hhEmap}--\ref{Fig4_hk0map0meV}. The ambient neutron background (BG) of 9 counts per minute was measured with Cd-blocked analyzer entrance and subtracted from all data. {\bQ}-independent BG arising from incoherent elastic scattering in the sample subtracted in Figure \ref{Fig4_hk0map0meV} was estimated by averaging $< 1\%$ of lowest-intensity points.

The experimental signatures of a liquid arise from well-defined microscopic local structures, fluctuating in time, that balance the energy of inter-molecular interactions and the entropy of thermal motion, which are at the origin of liquid state \cite{Poole_etal_Science1997,Poole_etal_Nature1992,MishimaStanley_Nature1998,Mallamace_PNAS2009,Huang_etal_PNAS2009,Clark_etal_PNAS2010,Beye_etal_PNAS2010}. In a glass, another common amorphous state of condensed matter, which is solid, the local order is static and survives averaging over practically infinite time, giving rise to elastic scattering. In a liquid it is dynamic, so a well-defined local structure is only present as an instantaneous snapshot. The corresponding equal-time correlation function is obtained by energy-integrating the inelastic scattering signal measured in neutron scattering experiment, where energy- and wave-vector-dependent patterns in Fourier space arising from time- and space-dependent real-space correlations are detected \cite{Squires}. For a system of electronic spins in a crystal, it is the correlation of magnetization fluctuations, rather than of atomic positions, that determines the character of the equilibrium state. The hallmark of spin liquid is the presence in magnetic neutron scattering intensity, in a broad range of energies, of a well-defined diffuse pattern in the wave vector space, which is characteristic of a particular short-range order. Indeed, if such a pattern persists in the energy range containing the dominant portion of scattering spectral weight, then this same pattern will dominate the single-time correlation describing the instantaneous local spin structure of the liquid state.

A liquid-like, diffuse scattering pattern persisting in a wide energy range, appears as a broad vertical column of intensity in the $(\bQ,E)$ coordinate plane cutting the 4-dimensional $(\bQ,E)$ phase space along particular wave vector trajectory \cite{Han_etal_Nature2012}. This is precisely how the measured intensity is shaped in FeTe$_{0.87}$S$_{0.13}$, Figure \ref{Fig1_hhEmap}. In Figure \ref{Fig1_hhEmap}, a--c, the $(\bQ,E)$ slice along the $(h,1-h)$ line of the two-dimensional (2D) Brillouin zone corresponding to the $a-b$ plane of Fe square lattice is shown, as it evolves upon cooling from 300 K to 3 K. At $h = 0.5$ this trajectory passes through $\bQ = (\pi,\pi)$, the point where spin resonance at $E \approx 7$ meV develops in the superconducting FeTe$_{1-x}$Se$_x$ samples \cite{Liu_etal_NatureMater2010,Lumsden_etal_NaturePhys2010}. We observe no resonance even at 3 K, which is below the superconducting transition temperature T$_c\approx 8$ K of our material \cite{Hu_etal_PRB2009}. This is consistent with the filamentary character of superconductivity \cite{Hu_etal_PRB2009}, which leaves bulk of the sample probed in our neutron experiment near critical doping and at the verge of being superconducting. The columns of magnetic scattering appear at $h \approx 0.3$ and $\approx 0.7$ in the units of P4/nmm (a = b $\approx 3.8$ \AA) reciprocal lattice, Figure \ref{Fig1_hhEmap}, d--f. This pattern is strikingly ubiquitous in FeTe$_{1-x}$Se$_x$ samples where superconductivity is partially or completely suppressed by imperfect composition, Fe substitution with Cu or Ni, or temperature \cite{Poole_etal_Science1997,Poole_etal_Nature1992,MishimaStanley_Nature1998,Mallamace_PNAS2009,Huang_etal_PNAS2009,Clark_etal_PNAS2010,Beye_etal_PNAS2010}.

The scattering we observe in FeTe$_{0.87}$S$_{0.13}$ does bear some similarity with parent Fe$_{1+y}$Te material: it is consistent with similarly large Fe local moment, which is also obtained from the Curie-Weiss analysis of magnetic susceptibility \cite{Hu_etal_PRB2009}, is stronger near $\bQ = (\pi,0)$, Figure \ref{Fig1_hhEmap}, g,h, and increases in intensity upon warming, which indicates thermally-enhanced electronic localization. However, there is also a marked distinction, which renders the four-spin up-up-up-up (UUUU) FM plaquette local structure of Figure \ref{Fig2_Plaquettes}a inapplicable. In Fe$_{1+y}$Te, where the overwhelming majority of magnetic spectral weight is also inelastic, spin-liquid-like, it can be very accurately described by the short-range AFM correlation of UUUU plaquettes illustrated in Figure \ref{Fig2_Plaquettes}, a,d \cite{Zaliznyak_etal_PRL2011}. The scattering intensity in this model is governed by the UUUU plaquette structure factor, which is exactly zero along the $(h,1-h)$ line and its square-symmetric replica, giving rise to zero-intensity square in Figure \ref{Fig2_Plaquettes}, d. Hence, while the practical absence of scattering along these lines in Fe$_{1+y}$Te provides unique fingerprint of the FM plaquette liquid, this model fundamentally cannot account for the intensity observed in FeTe$_{0.87}$S$_{0.13}$, Figure \ref{Fig1_hhEmap}, a--c, \cite{Zaliznyak_etal_PRL2011}.

The spatial structure of short-range spin-liquid correlation is revealed by constant-energy maps of magnetic scattering in Figs. \ref{Fig3_hk0map6meV} and \ref{Fig4_hk0map0meV}. The first important observation here is that all scattering features, including the characteristic ridges, which extend diagonally from $(\pm 0.5,0)$ and ($0,\pm 0.5)$ positions and correspond to diffuse columns in Figure \ref{Fig1_hhEmap}, a--c, are extremely broad. Resolution-corrected Lorentzian fits to the data shown in Figure \ref{Fig1_hhEmap}, d--h, suggest correlation length of only about one lattice repeat, which is the distance between the diagonal next-nearest neighbors Fe. Hence, the elementary local plaquette describing such correlation cannot consist of more than four Fe sites; the calculated structure factor for an ordered cluster of 8 spins already has features that are sharper than those seen in Figures \ref{Fig3_hk0map6meV} and \ref{Fig4_hk0map0meV}. Among all 4-spin plaquettes, the best fit to our data is provided by the slanted AFM up-down-up-down (UDUD) paquette shown in Figure \ref{Fig2_Plaquettes}, b,e. In the hindsight, the presence of this particular local order is quite natural: its long-range wave-like propagation with wave vector $(\pi,0)$ describes bicollinear AFM partial order observed in Fe$_{1+y}$Te, suggesting that the corresponding spin liquid state, Figure \ref{Fig2_Plaquettes}b, is close in energy.

In Fe$_{1+y}$Te parent material, the dominant UUUU plaquette correlation persists for temperatures up to at least T$_r = 300$ K, and only increases in intensity upon warming, as a result of thermally-induced localization of weakly itinerant electrons (the  effective spin localized on each Fe atom changes from S = 1 at T = 10 K to S = 3/2 at 300 K). That this correlation is so insensitive to thermal disorder suggests that formation of the UUUU plaquette spin liquid in Fe$_{1+y}$Te is governed by the hybridization energy notably exceeding k$_B$T$_r \approx 30$ meV. It remained a puzzle, why partial bicollinear magnetic order observed in Fe$_{1+y}$Te below T$_N \approx 60$ K, although it affects only minority of the spectral weight, does not accommodate UUUU plaquettes, thus indicating a different liquid state.  One solution to this puzzle is offered by the (co-)existence of two competing spin liquid polymorphs, only one of which crystallizes into an ordered structure on cooling. Our present results indeed provide strong experimental support to this idea. A fraction of magnetic spectral weight in FeTe$_{0.87}$S$_{0.13}$ also freezes in a short-range glassy state below $\approx 50$ K, corroborating that UDUD plaquettes are favored by energy rather than entropy, which favors higher-symmetry UUUU plaquettes.

Surprisingly, the ridges characteristic of the AFM UDUD plaquettes in FeTe$_{0.87}$S$_{0.13}$ become less pronounced upon warming, Figure \ref{Fig3_hk0map6meV}, b,c, while scattering intensity shifts to smaller Q, acquiring shape reminiscent of the UUUU plaquette liquid in Fe$_{1+y}$Te, Figure \ref{Fig2_Plaquettes}d. Similar temperature evolution is seen even more clearly in the pattern of (quasi-)elastic scattering, Figure \ref{Fig4_hk0map0meV}, which exhibits characteristic square pattern already at 100 K. We therefore introduce a model where the two contributions coexist, and refine the relative composition of two liquid components by fitting the data at different temperatures. While this model is not very sensitive to small admixtures of one of the components, it clearly shows that that UUUU plaquettes form a minority at 3 K, but a majority at 300 K. That such a simple model, with only few parameters, provides such an accurate description of all experimental data (see also Supporting Information) is a strong evidence that it correctly captures the physics of the observed spin liquid state, which is a mixture of two components with different local structure.

A closer inspection of scattering pattern in the UDUD plaquette model as a function of the inter-plaquette correlation length, $\xi$, leads to another surprising discovery. Recently, a variety of seemingly distinct, rather mysterious and ill-understood neutron patterns, including the spin resonance, were observed by different groups in various FeTe$_{1-x}$Se$_x$ samples, both with suppressed and well-developed superconductivity \cite{Liu_etal_NatureMater2010,Lumsden_etal_NaturePhys2010,Argyriou_etal_JPCM2010,Babkevich_etal_JPCM2012,Li_etal_PRL2010,Chi_etal_PRB2011,Babkevich_etal_PRB2011,Prokes_etal_PRB2012,Tsyrulin_etal_NJP2012,Xu_etal_PRL2012,Wen_etal_PRB2013}. In the absence of even a qualitative description, these results were often compared with ad-hoc functions, such as a Sato-Maki function, leading to data parameterizations void of much physical meaning \cite{Lumsden_etal_NaturePhys2010}. Remarkably, it appears that all the variety of the observed neutron patterns can be qualitatively well described by the UDUD plaqutte model with only slightly varying inter-plaquette correlation length, $\xi \sim 1-2$ lattice units. Hence, they all manifest very similar local correlation, revealing amazing intrinsic universality uncovered by our real-space model (see Supporting Information).

Thus, we find that the UDUD plaquette spin liquid emerges with doping in FeTe$_{1-x}$(S,Se)$_x$, as it begins to develop superconductivity. It breaks the C$_4$ symmetry of square lattice, and therefore can be related to "nematicity" observed in many unconventional superconductors \cite{Scalapino_RMP2012}. It is favored energetically, and, in FeTe is stabilized by bicollinear static antiferromagnetism. Symmetric UUUU plaquettes, on the other hand, are favored by entropy and are preferred at high temperature, as we observe.
Our findings suggest that a liquid-liquid phase transition occurs in the system of electronic spins in FeTe when it is doped, with Se or S, to become a superconductor. The newly emerging slanted UDUD plaquettes in all probability reflect a new electronic hybridization pattern, which is likely favored by the shift of atomic energy levels that results from doping, and which facilitates electron pairing. The structure of a plaquette is thus reflective of particular electronic Wannier functions, which define the orbital content of electronic bands in the tight binding view. That this change of the hybridization pattern belongs to the low-energy domain such that it is affected not only by small amounts of dopant atoms, but also by temperature, is a surprising discovery that calls for a profound revision of the tight binding paradigm. And so is the remarkable competition of two different electronic spin liquid polymorphs in a material on the threshold of unconventional superconductivity, which suggests new understanding of electronic nematicity and non-Fermi-liquid behavior \cite{Scalapino_RMP2012}.

\begin{acknowledgments}
We acknowledge discussions with J. Tranquada, W. Ku, G. Xu and B. Ocko. Work at BNL was supported by the Materials Sciences and Engineering Division, Office of Basic Energy Sciences, U.S. DOE under Contract No. DE-AC02-98CH10886. Research conducted at ORNL's High Flux Isotope Reactor and Spallation Neutron Source was sponsored by the Scientific User Facilities Division, Office of Basic Energy Sciences, U.S. Department of Energy.
\end{acknowledgments}


\begin{thebibliography}{40}

\bibitem{Mendeleev1887}
D.~Mendeléeff, {\em The compounds of ethyl alcohol with water},
J. Chem. Soc. Trans., \textbf{51}, 778--782 (1887).%

\bibitem{Rontgen1891}
W.~C.~R\"{o}ntgen, {\em \"{U}ber die Constitution des flüssigen Wassers},
Ann. d. Phys. u. Chem. N. F., \textbf{XLV}, 91--97 (1891).%

\bibitem{Pauling1939}
L.~Pauling, The Nature of Chemical Bond (Cornell University Press, Ithaca, New York 1939).

\bibitem{Poole_etal_Science1997}
P.~H.~Poole, T.~Grande, C.~A.~Angell, P.~F.~McMillan, %
{\em Polymorphic Phase Transitions in Liquids and Glasses},
Science \textbf{275}, 322--323 (1997).%

\bibitem{Poole_etal_Nature1992}
P.~H.~Poole, F.~Sciortino, U.~Essmann, H.~E.~Stanley,
{\em Phase behavior of metastable water},
Nature \textbf{360}, 324--328 (1992).%

\bibitem{MishimaStanley_Nature1998}
O.~Mishima, H.~E.~Stanley, %
{\em The relationship between liquid, supercooled and glassy water},
Nature \textbf{396}, 329--335 (1998).%

\bibitem{Mallamace_PNAS2009}
F.~Mallamace, %
{\em The liquid water polymorphism},
Proc. Natl Acad. Sci. USA \textbf{106}, 15097--15098 (2009).%

\bibitem{Huang_etal_PNAS2009}
C.~Huang, K.~T.~Wikfeldt, T.~Tokushima, D.~Nordlund, Y.~Harada, U.~Bergmann, M.~Niebuhr, T.~M.~Weiss, Y.~Horikawa, M.~Leetmaa, M.~P.~Ljungberg, O.~Takahashi, A.~Lenz, L.~Ojam\"{a}eg, A.~P.~Lyubartsev, S.~Shinc, L.~G.~M.~Pettersson, A.~Nilsson, %
{\em The inhomogeneous structure of water at ambient conditions},
Proc. Natl Acad. Sci. USA \textbf{106}, 15214--15218 (2009).%

\bibitem{Clark_etal_PNAS2010}
G.~N.~I.~Clark, G.~L.~Hura, J.~Teixeira, A.~K.~Soper, T.~Head-Gordon, %
{\em Small-angle scattering and the structure of ambient liquid water},
Proc. Natl Acad. Sci. USA \textbf{107}, 14003--14007 2010.%

\bibitem{Beye_etal_PNAS2010}
M.~Beye, F.~Sorgenfrei, W.~F.~Schlotter, W.~Wurth, A.~F\"{q}hlisch, %
{\em The liquid-liquid phase transition in silicon revealed by snapshots of valence electrons},
Proc. Natl Acad. Sci. USA \textbf{107}, 16772--16776 (2010).%

\bibitem{LeeBroholm_etal_Nature2002}
S.-H.~Lee, C.~Broholm, W.~Ratcliff, G.~Gasparovic, Q.~Huang, T.~H.~Kim, S.-W.~Cheong, %
{\em Emergent excitations in a geometrically frustrated magnet},
Nature \textbf{418}, 856--858 (2002).

\bibitem{StoneZaliznyak_etal_Nature2005}
M.~B.~Stone, I.~A.~Zaliznyak, T.~Hong, C.~L.~Broholm, D.~H.~Reich, %
{\em Quasiparticle breakdown in a quantum spin liquid},
Nature, \textbf{440}, 187--190 (2006).%

\bibitem{Zaliznyak_NatureMater2005}
I.~A.~Zaliznyak, %
{\em A glimpse of a Luttinger liquid},
Nature Mater. \textbf{4}, 273--275 (2005).%

\bibitem{Walters_etal_NaturePhys2009}
A. C. Walters, T. G. Perring, J.-S. Caux, A. T. Savici, G. D. Gu, C.-C. Lee, W. Ku, I. A. Zaliznyak, %
{\em Effect of covalent bonding on magnetism and the missing neutron intensity in copper oxide compounds.},
Nature Physics \textbf{5}, 867--872 (2009).%

\bibitem{Han_etal_Nature2012}
T.-H. Han, J. S. Helton, S. Chu, D. G. Nocera, J. A. Rodriguez-Rivera, C. Broholm, Y. Lee, %
{\em Fractionalized excitations in the spin-liquid state of a kagome-lattice antiferromagnet},
Nature \textbf{492}, 406--410 (2012).%

\bibitem{Anderson_Science1987}
P.~W.~Anderson, %
{\em The resonating valence bond state in La2CuO4 and superconductivity},
Science \textbf{235}, 1196--1198 (1987).%

\bibitem{LeeNagaosaWen_RMP2006}
P.~A.~Lee, N.~Nagaosa, X.-G.~Wen, %
{\em Doping a Mott insulator: physics of high-temperature superconductivity},
Rev. Mod. Phys. \textbf{78}, 17--85 (2006).%

\bibitem{Scalapino_RMP2012}
D.~J.~Scalapino, %
{\em A common thread: The pairing interaction for unconventional superconductors},
Rev. Mod. Phys. \textbf{84}, 1383--1417 (2012).

\bibitem{Hu_etal_PRB2009}
Rongwei Hu, Emil~S.~Bozin, J.~B.~Warren, C.~Petrovic, %
{\em Superconductivity, magnetism, and stoichiometry of single crystals of Fe$_{1+y}$(Te$_{1-x}$Se$_x$)$_z$},
Phys. Rev. B \textbf{80}, 214514 (2009).

\bibitem{Zaliznyak_etal_PRL2011}
I.~A.~Zaliznyak, Z.~Xu, J.~M.~Tranquada, G.~Gu, A.~M.~Tsvelik, M.~B.~Stone, %
{\em Unconventional Temperature Enhanced Magnetism in Fe$_{1.1}$Te},
Phys. Rev. Lett. \textbf{107}, 216403 (2011).

\bibitem{Liu_etal_NatureMater2010}
T.~J.~Liu, J.~Hu, B.~Qian, D.~Fobes, Z.~Q.~Mao, W.~Bao, M.~Reehuis, S.~A.~J.~Kimber, K.~Proke\^{s}, S.~Matas, D.~N.~Argyriou, A.~Hiess, A.~Rotaru, H.~Pham, L.~Spinu, Y.~Qiu, V.~Thampy, A.~T.~Savici, J.~A.~Rodriguez, C.~Broholm, %
{\em From ($\pi,0$) magnetic order to superconductivity with ($\pi,\pi$) magnetic resonance in Fe$_{1.02}$Te$_{1-x}$Se$_x$},
Nature Materials \textbf{9}, 717--720 (2010).

\bibitem{Lumsden_etal_NaturePhys2010}
M.~D.~Lumsden, A.~D.~Christianson, E.~A.~Goremychkin, S.~E.~Nagler, H.~A.~Mook, M.~B.~Stone, D.~L.~Abernathy, T.~Guidi, G.~J.~MacDougall, C.~de~la~Cruz, A.~S.~Sefat, M.~A.~McGuire, B.~C.~Sales, D.~Mandrus, %
{\em Evolution of spin excitations into the superconducting state in FeTe$_{1-x}$Se$_x$},
Nature Physics \textbf{6}, 182--186 (2010).

\bibitem{Argyriou_etal_JPCM2010}
D.~N.~Argyriou, A.~Hiess, A.~Akbari, I.~Eremin, M.~M.~Korshunov, J.~Hu, B.~Qian, Z.~Mao, Y.~Qiu, C.~Broholm, W.~Bao, %
{\em Incommensurate itinerant antiferromagnetic excitations and spin resonance in the FeTe$_{0.6}$Se$_{0.4}$ superconductor},
Phys. Rev. B \textbf{81}, 220503R (2010).

\bibitem{Babkevich_etal_JPCM2012}
P.~Babkevich, M.~Bendele, A.~T.~Boothroyd, K.~Conder, S.~N.~Gvasaliya, R.~Khasavov, E.~Pomjakushina, B.~Roessli, %
{\em Magnetic excitations of Fe$_{1+y}$Se$_x$Te$_{1-x}$ in magnetic and superconductive phases},
J. Phys.: Condens. Matter \textbf{22}, 142202 (2010).

\bibitem{Li_etal_PRL2010}
S.~Li, C.~Zhang, M.~Wang, H-Q.~Luo, X.~Lu, E.~Faulhaber, A.~Schneidewind, P.~Link, J.~Hu, T.~Xiang, P.~Dai, %
{\em Normal-State Hourglass Dispersion of the Spin Excitations in FeSe$_x$Te$_{1-x}$},
Phys. Rev. Lett. \textbf{105}, 157002 (2010).

\bibitem{Chi_etal_PRB2011}
S.~Chi, J.~A.~Rodriguez-Rivera, J.~W.~Lynn, C.~Zhang, D.~Phelan, D.~K.~Singh, R.~Paul, P.~Dai, %
{\em Common origin of the two types of magnetic fluctuations in iron chalcogenides},
Phys. Rev. B \textbf{84}, 214407 (2011).

\bibitem{Babkevich_etal_PRB2011}
P.~Babkevich, B.~Roessli, S.~N.~Gvasaliya, L.-P.~Regnault, P.~G.~Freeman, E.~Pomjakushina, K.~Conder, A.~T.~Boothroyd, %
{\em Spin anisotropy of the resonance peak in superconducting FeSe$_{0.5}$Te$_{0.5}$},
Phys. Rev. B \textbf{83}, 180506(R) (2011).

\bibitem{Prokes_etal_PRB2012}
K.~Prokes, A.~Hiess, W.~Bao, E.~Wheeler, S.~Landsgesell, D.~N.~Argyriou, %
{\em Anisotropy of the ($\pi,\pi$) dynamic susceptibility in magnetically ordered (x = 0.05) and superconducting (x = 0.40) Fe$_{1.02}$Te$_{1-x}$Se$_x$},
Phys. Rev. B \textbf{86}, 064503 (2012).

\bibitem{Tsyrulin_etal_NJP2012}
N.~Tsyrulin, R.~Viennois, E.~Giannini, M.~Boehm, M.~Jimenez-Ruiz, A.~A.~Omrani, B.~Dalla Piazza, H.~M.~Rønnow, %
{\em Magnetic hourglass dispersion and its relation to high-temperature superconductivity in iron-tuned Fe$_{1+y}$Te$_{0.7}$Se$_{0.3}$},
New Journ. Phys. \textbf{14}, 073025 (2012).

\bibitem{Xu_etal_PRL2012}
Z.~Xu, J.~Wen, Y.~Zhao, M.~Matsuda, W.~Ku, X.~Liu, G.~Gu, D.-H.~Lee, R.~J.~Birgeneau, J.~M.~Tranquada, G.~Xu %
{\em Temperature-Dependent Transformation of the Magnetic Excitation Spectrum on Approaching Superconductivity in Fe$_{1+y-x}$(Ni/Cu)$_x$Te$_{0.5}$Se$_{0.5}$},
Phys. Rev. Lett. \textbf{109}, 227002 (2012).

\bibitem{Wen_etal_PRB2013}
J.~Wen, S.~Li, Z.~Xu, C.~Zhang, M.~Matsuda, O.~Sobolev, J.~T.~Park, A.~D.~Christianson, E.~Bourret-Courchesne, Q.~Li, G.~Gu, D.-H.~Lee, J.~M.~Tranquada, G.~Xu, R.~J.~Birgeneau, %
{\em Enhanced low-energy magnetic excitations via suppression of the itinerancy in Fe$_{0.98-z}$Cu$_z$Te$_{0.5}$Se$_{0.5}$},
Phys. Rev. B \textbf{88}, 144509 (2013).

\bibitem{Squires}
G.~L.~Squires, Introduction to the Theory of Thermal Neutron Scattering (Cambridge University Press, 1978, 2012).


\end{thebibliography}
\end{document}